\documentclass[twocolumn]{aastex631}
\usepackage[utf8]{inputenc}
\usepackage{graphicx}
\usepackage{amsmath}
\usepackage{booktabs}
\usepackage{graphicx} 
\usepackage[flushleft]{threeparttable}

\shorttitle{Huang et al. reply}
\graphicspath{{./}{figures/}}

\begin{document}
\title{Reply to Comment on ``A slightly oblate dark matter halo revealed by a retrograde precessing Galactic disk warp"}

\author{Yang Huang}
\affiliation{School of Astronomy and Space Science, University of Chinese Academy of Sciences, Beijing, 100049, China}
\affiliation{National Astronomical Observatories, Chinese Academy of Science, Beijing, 100012, China}
\author{Qikang Feng}
\affiliation{Department of Astronomy Peking University, Beijing, 100871, China}
\affiliation{Kavli Institute for Astronomy and Astrophysics, Peking University, Beijing, 100871, China}
\author{Tigran Khachaturyants}
\affiliation{Department of Astronomy, School of Physics and Astronomy, Shanghai Jiao Tong University, Shanghai 200240, China}
\affiliation{Key Laboratory for Particle Astrophysics and Cosmology (MOE)/Shanghai Key Laboratory for Particle Physics and Cosmology, Shanghai 200240, China}
\author{Huawei Zhang}
\affiliation{Department of Astronomy Peking University, 
Beijing, 100871, China}
\affiliation{Kavli Institute for Astronomy and Astrophysics, Peking University, Beijing, 100871, China}
\author{Jifeng Liu}
\affiliation{National Astronomical Observatories, Chinese Academy of Science, Beijing, 100012, China}
\affiliation{New Cornerstone Science Laboratory, National Astronomical Observatories, Chinese Academy of Sciences, Beijing 100012, China}
\affiliation{School of Astronomy and Space Science, University of Chinese Academy of Sciences, Beijing, 100049, China}
\author{Juntai Shen}
\affiliation{Department of Astronomy, School of Physics and Astronomy, Shanghai Jiao Tong University, Shanghai 200240, China}
\affiliation{Key Laboratory for Particle Astrophysics and Cosmology (MOE)/Shanghai Key Laboratory for Particle Physics and Cosmology, Shanghai 200240, China}
\author{Timothy C. Beers}
\affiliation{Department of Physics and Astronomy and JINA Center for the Evolution of the Elements (JINA-CEE), University of Notre Dame, Notre Dame, IN 46556, USA}
\author{Youjun Lu}
\affiliation{National Astronomical Observatories, Chinese Academy of Science, Beijing, 100012, China}
\affiliation{School of Astronomy and Space Science, University of Chinese Academy of Sciences, Beijing, 100049, China}
\author{Song Wang}
\affiliation{National Astronomical Observatories, Chinese Academy of Science, Beijing, 100012, China}
\author{Haibo Yuan}
\affiliation{Department of Astronomy, Beijing Normal University, Beijing, 100875, People's Republic of China}

\begin{abstract}
In this reply, we present a comprehensive analysis addressing the concerns raised by \cite{D24} regarding our recent measurement of the disk warp precession using the `motion-picture' method \citep{H24}.
We carefully examine the impact of ignoring the twist of the disk warp and the so-called $R$-$\tau$ correlation on the estimation of the precession rate. 
The results indicate that the effect is minor and does not exceed the systematic and statistical uncertainties.
Using N-body+SPH simulation data, we confirm that the `motion-picture' technique is effective in measuring retrograde precession of disk warp in stellar populations younger than 170 Myr, similar to classical Cepheids. 
Therefore, the overall conclusions of \cite{H24} remain robust.
\end{abstract}

\section{Main}
In \cite{H24}, we measured the precession of the Galactic disc warp using a `motion-picture' method, by tracing the line-of-nodes (LON) angle, $\phi_{w}$, of classical Cepheids as a function of their age, $\tau$. 
By analyzing a sample of 2,600 classical Cepheids with accurately determined distances and ages, we found that the Galactic warp is undergoing mild retrograde precession at a rate of $-2.1 \pm 0.5 ~({\rm statistical}) \pm 0.6~({\rm systematic})$ km~s$^{-1}$~kpc$^{-1}$.
This result is further used to constrain the shape of the inner dark matter halo, suggesting it is slightly oblate, that is required to explain the measured mild retrograde precession.

\cite{D24} raised concerns that: 1) the `motion-picture' method is ineffective, as young stars like classical Cepheids primarily trace the gas warp at present, i.e. $\phi_{w}$ should hardly depend on Cepheid age; and 2) the measured d$\phi_{w}/$d$\tau$ is mainly from an omitted-variable bias, resulting from neglecting the natural twist of the disc warp, d$\phi_{w}/$d$R$, and the relationship between $R$ and $\tau$ for classical Cepheids.

As demonstrated in earlier $N$-body + smooth particle hydrodynamics ($N$-body+SPH) simulations \citep{TK21}, young mono-age stellar populations that were born in the warp experience orbital tilting and phase-mixing over much longer timescales than in the `motion-picture' method: at least 300 Myr are needed for major changes to manifest in these populations. 
A detailed analysis of a warped simulation from \cite{TK22} shows that the retrograde precession of the disc warp can be revealed by the  `motion-picture' method, using stellar populations with ages younger than 170\,Myr (similar to classical Cepheids).
Although not discussed in \cite{H24}, we find that neglecting the twist of the disc warp has only a minor effect on the measured d$\phi_{w}/$d$\tau$, contributing no more than the reported systematic uncertainty from other sources.
In summary, the main findings of \cite{H24} remain valid.

\begin{figure*}
	\begin{center}
    \includegraphics[width=1\textwidth]{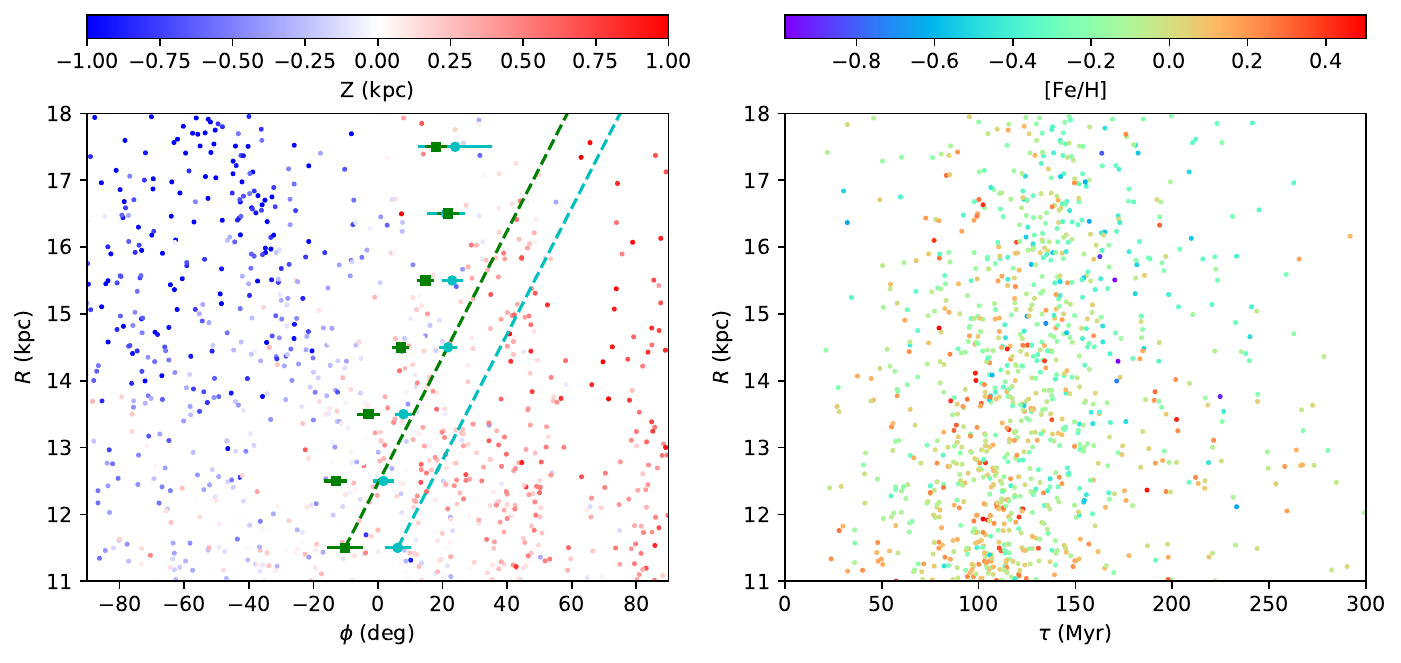}
	\end{center}
	\caption{Left panel: The distribution of Cepheids sample over Galactic radius $R$ and azimuth $\phi$ (increasing in the direction of Galactic rotation), color-coded by $Z$ (similar to Figure~1 in \cite{D24}). Here, we add the LON measurements at each radius on the plot. The green squares represent the measurements by our sample and the cyan dots denote the measurements by \cite{CH19}. Two dashed lines show the slope of d$\phi_{w}/$d$R$=$10.6$~deg~kpc$^{-1}$ reported by \citet{D24}, starting from the first point. It is clear that the slope suggested by \cite{D24} deviates from our actual data, which exhibits a much shallower and more complex trend (see text for details).
     Right panel: The distribution of the Cepheids sample over Galactic radius $R$ and age $\tau$, color-coded by metallicity [Fe/H]. The Pearson correlation coefficient for the stars in this plot is only 0.16.}
	\label{Fig1}
\end{figure*}

\begin{table*}
    \centering
    \begin{threeparttable}
    \caption{The Effect of an Omitted-variable Bias for Different Radial Bins} \label{Table1}
    \begin{tabular*}{\textwidth}{@{\extracolsep{\fill}}cccccc}
    \hline
    Bin & d$\phi_{w}/$d$R$& $\beta_{\tau}^{R}$ & d$\phi_{w}/$d$\tau$\tnote{a}\\
    &(degrees~kpc$^{-1}$)&(degrees~Myr$^{-1}$)&(degrees~Myr$^{-1}$)\\
    \hline
    All ($R>7.5\ \mathrm{kpc}$) & 4.4$\pm$0.7 & 0.03$\pm$0.01& 0.12$\pm$0.03 (stat)$\pm 0.04$ (sys) \\
    $11.8<R<18.8\ \mathrm{kpc}$ & 6.1$\pm$1.1 &0.04$\pm$0.01 & 0.12$\pm$0.01 (stat)$\pm 0.04$ (sys)\\
    $14.0<R<21.0\ \mathrm{kpc}$ &3.9$\pm$0.8 &0.03$\pm$0.01 & 0.09$\pm$ 0.04 (stat)$\pm 0.04$ (sys)\\
    $R>15.5\ \mathrm{kpc}$ &2.1$\pm$1.8 &0.01$\pm$0.01 & 0.07$\pm$0.02 (stat)$\pm 0.04$ (sys) \\
    $9.0 < R < 12.0\ \mathrm{kpc}$ &$-3.9\pm$6.9& $-0.03\pm$0.05 &0.07$\pm$0.33 (stat)$\pm 0.04$ (sys)\\
    \hline
    \end{tabular*}
    \begin{tablenotes}
    \footnotesize
    \item[a] Taken from \citet{H24}.
    \end{tablenotes}
    \end{threeparttable}
\end{table*}

\section{The effect of an omitted-variable bias}
\cite{D24} argued that the positive value of d$\phi_{w}/$d$\tau$ measured in \cite{H24} is primarily due to an omitted-variable bias, resulting from the neglect of the natural twist d$\phi_{w}/$d$R$ and the correlation between $R$ and $\tau$ for classical Cepheids.
However, they overestimated both the value of d$\phi_{w}/$d$R$ and the correlation coefficient between $R$ and $\tau$.
For the twist term, the LON $\phi_{w}$ of the disc warp does not increase monotonically with $R$, which is different from the trend reported in \cite{D24}. 
Both our Figure~1 here and Figure~3 of \cite{CH19} clearly show that $\phi_{w}$ initially decreases with $R$ from the warp's starting radius to $R \sim 12-13$~kpc, then increases with $R$ until $R \sim 15-16$~kpc, and finally tends toward being flat.
The slope of d$\phi_{w}/$d$R$ for all sample stars with $R > 7.5$~kpc is only $4.4 \pm 0.7$~deg~kpc$^{-1}$. 
Specifically, for the three radial bins analyzed in \cite{H24}, the slopes are $6.1 \pm 1.1$~deg~kpc$^{-1}$ for $11.8 < R < 18.8$~kpc, $3.9 \pm 0.8$~deg~kpc$^{-1}$ for $14.0 < R < 21.0$~kpc, and $2.1 \pm 1.8$~deg~kpc$^{-1}$ for $R > 15.5$~kpc (see Table~1). All of these detected slopes are significantly smaller than the value of $10.6 \pm 0.8$~deg~kpc$^{-1}$ reported by \cite{D24}.

\cite{D24} also claimed that a correlation exists between $R$ and $\tau$ for classical Cepheids, which they attribute to the metallicity gradient of the Galactic disc and the Cepheid age–metallicity relation. However, as discussed in \cite{H24}, the contribution of the metallicity term to age estimation for Cepheids is negligible compared to the period term \citep{DS21}. \cite{H24} demonstrated that assuming a constant [Fe/H] value of $-0.07$ for all stars does not affect the calculated ages of classical Cepheids. Consequently,  we do not expect a significant $R$-$\tau$ correlation for Cepheids, given the weak age–metallicity relation. As a verification, we calculated the Pearson correlation coefficient $\rho_{R\tau}$ for the stars in Figure~1, which is only 0.16, consistent with our expectations.

We then apply Equation~6 from \cite{D24} to assess the impact of the disc-warp twist and the $R$-$\tau$ correlation (although this is scarcely a true correlation) on the estimated d$\phi_{w}/$d$\tau$ as reported by \cite{H24}.
The details are summarized in Table~1.
For all sample stars, this effect is just $0.03 \pm 0.01$~deg~Myr$^{-1}$, only one-third of the value claimed by \cite{D24}, and comparable to the systematic or statistical uncertainties in the estimation of d$\phi_{w}/$d$\tau$ reported by \cite{H24}.
Similar results are also found for the three radial bins: $11.8 < R < 18.8$~kpc, $14.0 < R < 21.0$~kpc, and $R > 15.5$~kpc.
Moreover, if the argument of \cite{D24} is correct, we should expect a negative value of d$\phi_{w}/$d$\tau$ for $9 < R < 12$~kpc, as $\phi_{w}$ decreases with $R$ within this radial range.
However, our measurements show a positive value (despite the relatively large error; see Table~1), which serves as a counterargument.

\section{Testing `motion-picture' method with simulations}
Here we use simulation data to demonstrate that: 1) young stellar populations preserve the orbital information imparted by the gas at the time of their formation; and 2) applying the `motion-picture' method to these young stellar populations can reveal a retrograde precession.

\begin{figure}
	\begin{center}
    \includegraphics[width=0.465\textwidth]{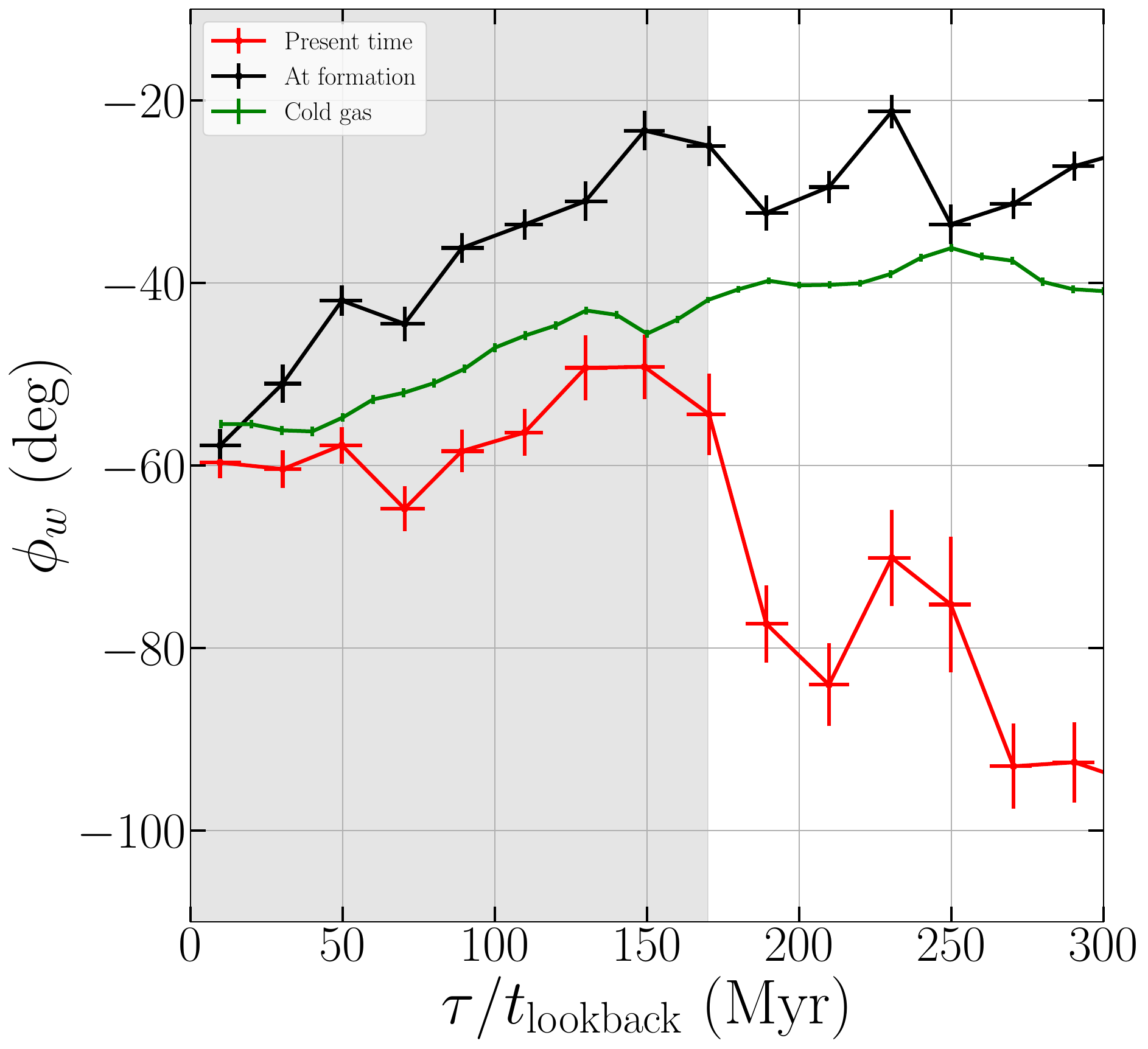}
	\end{center}
	\caption{
                 Similar to Fig.~1(d) of \cite{H24}, the measured LON, as a function of median age for different bins of young stellar populations, from one time period in an $N$-body+SPH simulation \citep{TK21,TK22}. 
                 Here, age is determined as the time difference between the formation time of the stars and the end of a given time period (i.e., the present time); relative lookback time, $t_{\rm lookback}$, represents the time elapsed prior to the present time. 
                 The LON of the cold gas warp is shown as a green solid line and clearly exhibits retrograde precession. The LON of young stellar populations are presented both at the time of formation (black solid line) and at present time (red solid line). 
                 The overall offset between gas and stars at formation is due to the majority of older stars currently located at $12 < R < 13$~kpc forming outside this annulus; the two curves overlap when only stars born in the annulus are selected.}
	\label{Fig1-1}
\end{figure}

We adopt a Milky-Way-like warped $N$-body+SPH simulation described in \citealt{TK22} (henceforth KH22), which is evolved using the code \textsc{gasoline} \citep{Wadsley+04} for $12\;\rm{Gyr}$ without any mergers. In this warped model, the dark matter halo is triaxal with the angular momentum of the embedded gas corona misaligned with its principal axes; this results in gas accreting via an S-shaped warp onto the disc. Detailed discussion of the initial conditions and evolution of the warped model is presented in KH22.  In each snapshot of the simulation, similar to the pre-processing in KH22, the stellar disc is centred and rotated into the ($x,y$) plane based on the angular momentum of the inner region ($R<5$~kpc). As a result of the pre-processing, the sense of rotation in the model is in the direction of increasing azimuth.
The warp in the KH22 simulation is long-lived, and increases in size with time; it has been shown in K22 (via spectral analysis) to have a mostly static or mildly retrograde precessing warp.
We then select several hundred time periods between $4$ Gyr and $12$ Gyr that have clear disc warp precession. These time periods are determined by analysing the LON of the cold gas ($\rm{T}\leq50,000\,\rm{K}$) at a fixed radial annulus $12 < R <13$~kpc.
Here we define LON as the $\phi_w$ angle of the radially averaged angular momentum vector.
For each time period, we then analyse the LON of young stellar populations separated into equally-sized, non-overlapping age bins ($\Delta\tau=20\,\rm{Myr}$).

As an example, we present the LON for cold gas, stars at formation, and stars at present as functions of age or relative lookback time for one of these time periods in Figure~\ref{Fig1-1}. 
Clearly, the LON versus age of stars at present basically catches the trend of both cold gas and stars at formation for stellar populations younger than 170~Myr, which contradicts the argument of \cite{D24} that the `LON should hardly depend on Cepheid age'.
Discrepancies begin to emerge beyond this age, suggesting that the warp signal preserved from the gas at the time of formation starts to be erased.
Following the `motion-picture' approach in \cite{H24}, we perform linear fits to the LON as a function of median age for two age intervals: [0,110]~Myr and [0,150]~Myr.
Here, 110 Myr and 150 Myr represent the median and upper age limits, respectively, of the classical Cepheids sample in \cite{H24}.
For all picked time periods, we find that 90\% and 80\% of the disc warp measured with retrograde precession using the `motion-picture' method, for the [0,110] Myr and [0,150] Myr age bins respectively, are indeed retrograde, as indicated by the stars at the time of formation.
A more thorough validation and analysis of the `motion-picture' method will be performed in a forthcoming paper.

These simulation results can also be understood physically as follows.
\cite{TK21} demonstrated that young stellar populations born in the warp will gradually tilt and phase-mix into the disc. However, these processes are slow and can take up to 300 Myr to produce significant changes within a given mono-age population.
Therefore, it is possible that there exists a maximum population age at which the warp history can be preserved: within a retrograde precessing warp, older populations are born with a larger LON, but their orbital precession is not sufficient enough to erase the warp signal until a population is older than about 170~Myr.
The above explanations provide a theoretical foundation for \cite{H24} in measuring disc warp precession using the `motion-picture' technique with classical Cepheids samples (mostly younger than 170 Myr).

Finally, we emphasize that: 1) the `motion-picture' method is effective for young stellar populations, particularly those younger than 170~Myr (such as the classical Cepheids used in \citealt{H24}), whose angular momentum has yet to be altered to such a degree that the warp signal were unrecognisable \citep{TK21}; and 2) neglecting the disc-warp twist has a minor effect on the estimated d$\phi_{w}/$d$\tau$, if any, and this effect is no greater than the systematic or statistical uncertainties. Therefore, the main conclusion of \cite{H24} remains valid.

\end{document}